\documentclass[sigconf,screen]{acmart}

\acmConference[ICSE 2024]{46th International Conference on Software Engineering}{April 2024}{Lisbon, Portugal}

\AtBeginDocument{%
  \providecommand\BibTeX{{%
    \normalfont B\kern-0.5em{\scshape i\kern-0.25em b}\kern-0.8em\TeX}}}

\setcopyright{acmcopyright}
\copyrightyear{2018}
\acmYear{2018}
\acmDOI{XXXXXXX.XXXXXXX}

\usepackage{dirtytalk}
\usepackage{wrapfig}
\usepackage{tikz}

\newcommand{\sys}{{\small{PADLOCK}}}

\newcommand{\add}[1]{\textcolor{black}{#1}}

\copyrightyear{2024}
\acmYear{2024}
\setcopyright{acmlicensed}\acmConference[IDE '24]{2024 First IDE
Workshop}{April 20, 2024}{Lisbon, Portugal}
\acmBooktitle{2024 First IDE Workshop (IDE '24), April 20, 2024, Lisbon,
Portugal}
\acmDOI{10.1145/3643796.3648453}
\acmISBN{979-8-4007-0580-9/24/04}

\begin{document}

\title{\add{``Don't Step on My Toes'': Resolving Editing Conflicts in Real-Time Collaboration in Computational Notebooks}}

\author{April Yi Wang}
\orcid{0000-0001-8724-4662}
\email{april.wang@inf.ethz.ch}
\affiliation{%
  \institution{ETH Zürich}
  \country{Switzerland}
}

\author{Zihan Wu}
\orcid{0000-0002-3161-2232}
\email{ziwu@umich.edu}
\affiliation{%
  \institution{University of Michigan}
  \country{USA}
}

\author{Christopher Brooks}
\orcid{0000-0003-0875-0204}
\email{brooksch@umich.edu}
\affiliation{%
  \institution{University of Michigan}
  \country{USA}
}

\author{Steve Oney}
\orcid{0000-0002-5823-1499}
\email{soney@umich.edu}
\affiliation{%
  \institution{University of Michigan}
    \country{USA}
}

\renewcommand{\shortauthors}{Wang et al.}

\begin{abstract}
Real-time collaborative editing in computational notebooks can improve the efficiency of teamwork for data scientists. 
However, working together through synchronous editing of notebooks introduces new challenges. 
Data scientists may inadvertently interfere with each others' work by altering the shared codebase and runtime state if they do not set up a social protocol for working together and monitoring their collaborators' progress. 
In this paper, we propose a real-time collaborative editing model for \add{resolving conflict edits} in computational notebooks that introduces three levels of edit protection to help collaborators avoid introducing errors to both the program source code and changes to the runtime state. 
\end{abstract}





\maketitle

\section{Introduction}
``You work on \emph{this} section, and I'll work on \emph{that} one'' is a familiar refrain for authors who work in teams.
Working on different parts of the same document is a natural way to  combine collaborators' work and avoid conflicts~\cite{posner1992people}.
In data science programming, collaborators use a variety of collaborative strategies including ``divide and conquer'' (splitting work between team members) and ``competitive authoring'' (working on the same sub-problem simultaneously)~\cite{wang2019data}.
However, Jupyter and other computational notebooks, which are often used by data scientists, introduce new challenges for collaboration.
Although some version control tools (e.g., Git) work for computational notebooks, they mostly support the collaboration strategies for dividing work (e.g., working in independent files).
Further, data scientists sometimes collaborate \emph{synchronously}, with tools like JupyterLab~\cite{jupyterlab}, Google Colab~\cite{googlecolab}, and Deepnote~\cite{deepnote} that broadcast code and runtime updates to collaborators in real-time~\cite{wang2019data}.

Synchronized collaborative computational notebooks allow data scientists to immediately share the notebook edits and the runtime state, which improves data science teamwork by creating a shared context, encouraging more explanation, reducing communication costs, and improving reproducibility \cite{wang2019data, kross2019practitioners}.
However, these synchronized notebooks also introduce many unique collaboration challenges~\cite{wang2019data}.
For example, one collaborator might inadvertently change the runtime state and indirectly break another collaborator's code in a way that is difficult to debug~\cite{wang2019data}.


Inspired by these challenges and opportunities, we propose a set of interactive techniques to minimize collaboration friction while maintaining the readability of the shared notebook.
We instantiate these techniques in \sys{}\footnote{An acronym: \textbf{P}arallelization \textbf{A}nd \textbf{D}ata \textbf{L}ocks \textbf{O}ffset \textbf{C}ollaboration \textbf{K}inks}, an extension to the open source JupyterLab platform.
\sys{} provides three domain-relevant mechanisms to improve collaboration on computational narratives.
\emph{Cell-level access control}, allows collaborators to control who can view or edit cells to better support common collaboration patterns.
\emph{Variable-level access control}, extends this access control from code to runtime values to prevent implicit editing conflicts.
\emph{Parallel cell groups}, allows collabors' edits to be scoped to allow them to pursue exploratory solutions independent of collaborators.
Our evaluation of \sys{} has shown that these mechanisms can effectively prevent editing conflicts in shared notebooks and they support a wide range of collaborative workflows.

This work makes several contributions that advance the state of the art for collaborative data science tools: three new mechanisms (cell-level access controll, variable-level access control, and parallel cell groups) to improve collaborative data science work; a system (\sys{}) that instantiates all these features in a JupyterLab plugin.

\begin{figure*}
\includegraphics[width=\textwidth]{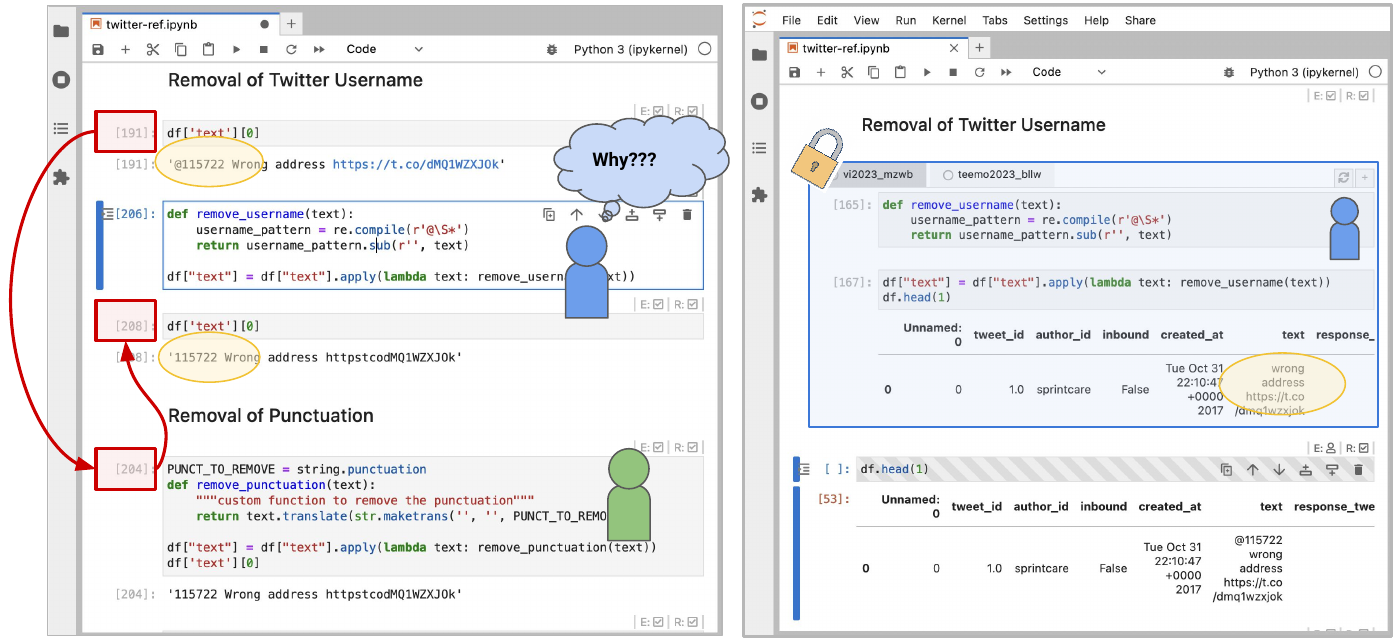}
  \caption{Editing conflicts in real-time collaborative notebooks can be implicit. As shown on the left, one can get an unexpected execution result because the collaborator accidentally changed the shared variable. As shown on the right, \sys{} helps data scientists \add{resolve editing conflicts} in real-time collaborative editing in computational notebooks. }
  \label{fig:teaser}
\end{figure*}

\noindent{}To the best of our knowledge, \sys{} is the first tool to:
\begin{itemize}
    \item Give users the ability to specify access control constraints at the level of individual cells in computational notebooks
    \item Allow programmers to specify which collaborators (as opposed to which code) can access or overwrite variables
    \item Allow data scientists to work in ``parallel cell groups'', that are scoped in a way where they can access and reference each other's work without worrying about introducing conflicting code.
\end{itemize}

\section{Background}

Data science programming can benefit from both synchronous and asynchronous collaboration.
Zhang et al. \cite{zhang2020data} conducted a large-scale survey of data science workers and found that data science work is \say{extremely collaborative} and tools greatly influence their collaboration practices.
Wang et al. \cite{wang2019data} found that compared to individual programming contexts, real-time collaboration can encourage more exploration and provide a shared context for communication.
They proposed four collaboration styles to characterize how data scientists work together, including \textit{single authoring} where one collaborator does the majority of the work, \textit{pair authoring} where one collaborator contributes to the implementation while the other collaborator participates in the discussion, \textit{divide and conquer} where collaborators divide the task into subgoals and assign to each other, and \textit{competitive authoring} where collaborators implement independently toward the same goal.

Researchers have proposed different systems to support programmers in working on the same code file synchronously. 
Targeting novice programmers, Warner and Guo created CodePilot \cite{warner2017codepilot}, which is the first real-time collaborative programming tool that embeds coding, testing, bug reporting, and VCS features. R\"{a}dle et al. created Codestrates \cite{radle2017codestrates} to embed literate computing based on a shareable dynamic media system \cite{klokmose2015webstrates} and enables users to collaboratively work on authoring and debugging.
There are also other tools provided in the form of IDE plugins \cite{saros, liveshare}. For example, Microsoft Live Share in VSCode \cite{liveshare} allows users to set the code to read-only for collaborators or enable server sharing for collaborating with the same variables.
On a more fine-grained level, some researchers focused on resolving editing conflict of collaborative real-time editing in rich text with Conflict-Free Replicated Data Types (CRDTs) \cite{litt2022peritext}; others examined collaboration in a broader context of peer assessment in programming classes for lightweight test cases \cite{wang2021puzzleme}.
Real-time collaboration in programming also brings unique challenges. Goldman \cite{goldman2012software} identified that syntax errors introduced by other collaborators might block one programmer's work. To address the issue, Goldman et al. proposed Collabode \cite{goldman2011real} which uses error-mediated integration that only integrate edits that do not cause compile errors.

For data science work, although there are many features that aims at improving awareness between collaborators and enhance communication, there has been limited features regarding preventing conflicts or interference in the collaboration process.
Although the Deepnote~\cite{deepnote} and Hex~\cite{hex} computational notebook platforms provide some support for preventing conflicts---for example, both can prevent simultaneous editing of the same cell---they do not give collaborators fine-grained control over how this works, as \sys{} does.
For example, both tools still allow edits after the ``cell owner'' stops editing the cell, even if they were only pausing their activity and planning to resume shortly thereafter.

\section{Design Motivations}

We describe three examples conflict-causing real-time collaboration scenarios in data science, based on challenges identified in prior work~\cite{wang2019data}.

\textbf{Simultaneous Feature Implementation:}
Conflicts may arise during ``competitive authoring''~\cite{wang2019data}, where data scientists work on the same problems simultaneously.
To avoid conflicts on interdependent code, collaborators must either be in close communication or edit their own copies of the shared code and coordinate an eventual merge.
This can be prohibitively difficult, particularly in large teams.

\textbf{Concurrent Variable Use or Modification:}
Even if they are not working on the same features, collaborators often need to work on the same shared data.
For example, when collaborators work on a shared dataframe, it can be easy to accidentally make edits that conflict with their collaborators' work.
However, there are still also occasions where data scientists need to synchronize changes made by their collaborators working upstream.
Coordinating work on these shared variables can be difficult and error-prone.
%

\textbf{Social Concerns in Real-Time Collaboration:}
Prior work has found that authors have social concerns about letting collaborators see their intermediate work~\cite{wang2017users}.
For example, they might fear being judged about the quality of their intermediate code and novices may be self-conscious about their visible progress.
In these scenarios, collaborators may want to take steps to gain some degree of privacy as they work.
Normally, this must be done by working in a non-shared document and merging their work when they deem that it is `ready' to be integrated~\cite{wang2017users}.
However, this strategy can be difficult to implement in many scenarios where collaborators' work is interdependent, as is frequently the case~\cite{wang2019data}.
\begin{figure}
\begin{center}
\includegraphics[width=1.0\linewidth]{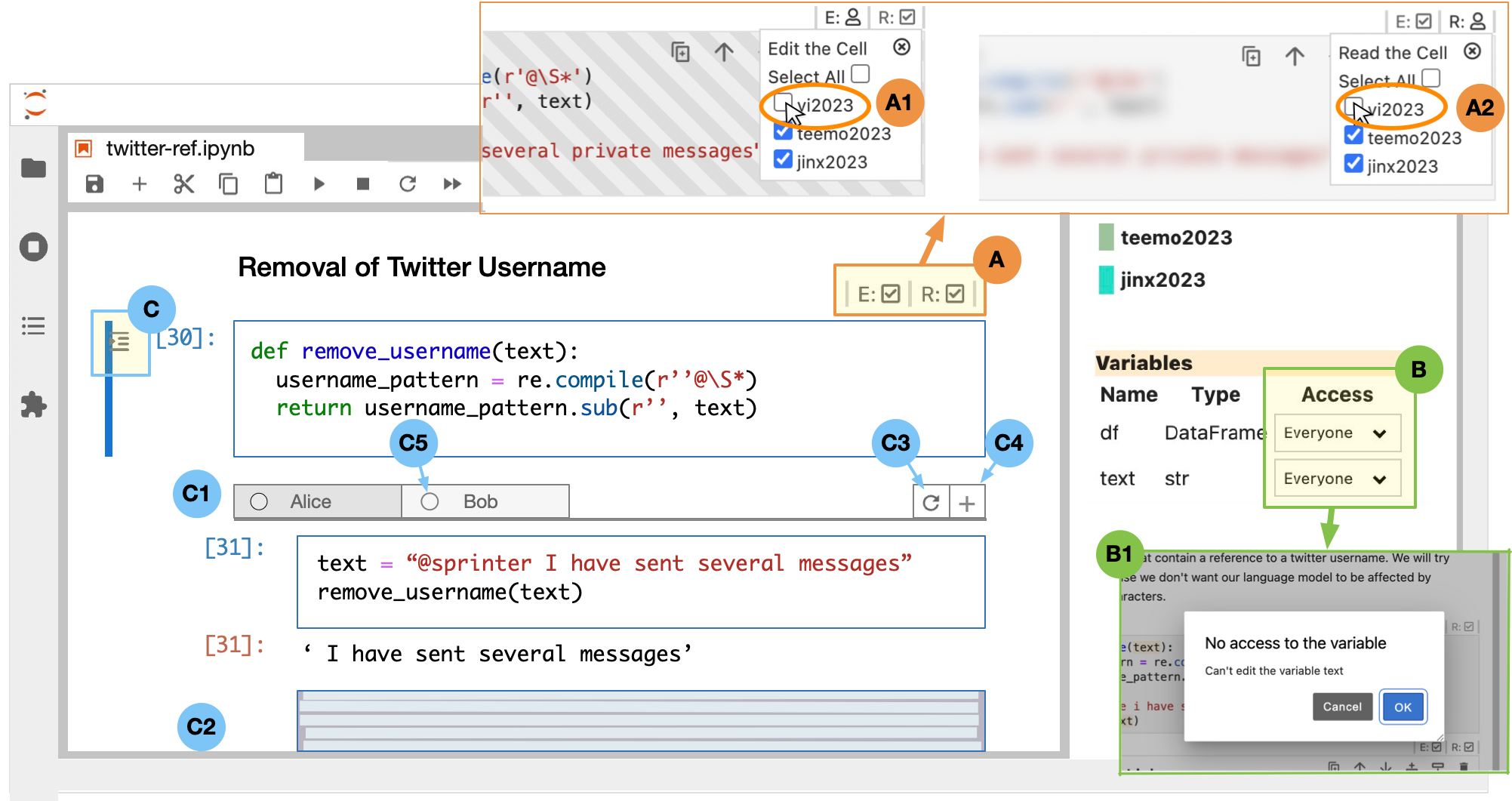}
\end{center}
  \vspace{-1em}
  \caption{Overview of the three conflict-free mechanisms in \sys{}. (A) Cell-level access control allows collaborators to claim ownership of the code cells and restrict others from editing or viewing them: (A1) Unchecking edit access for a user will change the cell background and disable editing; (A2) Unchecking read access for a user will blur the cell. (B) Variable-level access control extends the idea of access control from cells to shared variables: (B1) Unchecking a user's variable access; (B2) After losing access, they will not be able to edit the variable and will receive warning when attempting to do so. (C) Parallel cell groups define a designated area where changes of the code and runtime state stay inside its own scope: (C1) One parallel cell group can contain multiple tabs; (C2) Each tab can contain multiple cells; (C3) Sync the variables from the global scope to the current active tab; (C4) Add a new tab; (C5) Click the radio button to mark it as the ``main'' tab.}
  \label{fig:overview}
  \vspace{-2em}
\end{figure}
\section{System Overview}


\subsection{Cell-Level Access Control}
Computational notebooks consist of \emph{cells}.
Each cell typically represents a conceptual unit within the larger notebook.
For example, a notebook might consist of one cell to fetch data from a remote API, another to clean those data, and other cells for various transformations and visualizations of the data.
In \sys{}, we leverage the structure of cells in order to allow collaborators to claim ownership of parts of larger collaborative notebooks.
This helps address of the challenges of synchronous editing in traditional text-based programming tools where there are no clear ``dividing lines'' between different parts of the shared codebase and unclear how to localize the scope and effects of collaborators' changes.

Specifically, \sys{} enables \emph{cell-level access control} where users can prevent collaborators from viewing or editing a collection of cells.
As Figure~\ref{fig:overview}.A shows, users can select a code cell and specify who can read or edit the code.
As prior studies have found, there are many collaboration styles~\cite{wang2019data}, and cell-level access control benefits multiple collaboration styles.
In a ``single authoring'' style~\cite{wang2019data}---where one collaborator contributes the majority of ideas and code---setting cells to be only editable by the main contributor can prevent others from accidentally introducing errors.
In a ``divide and conquer'' style~\cite{wang2019data}---where collaborators split up work---restricting view access might ease feelings of self-consciousness that authors sometimes feel when collaborators can see their writing in real-time (which might otherwise lead them to work in a private editor and then copy its contents to the main notebook, as prior work found in collaborative writing~\cite{wang2017users}).

When an author is restricted from editing a cell, the background of the cell (grey striping) indicates that edit access is not permitted.
When an author is restricted from reading a cell, the content of the cell is blurred but activity (and thus awareness of contributors' location in the narrative) is supported.
%
Thus, the view control of the cell can allow them to focus on early explorations of ideas while still letting others know what they are working on.

\subsection{Variable-Level Access Control}
Restricting cell access gives collaborators control of code edits but it does not prevent collaborators from modifying the shared runtime state.
For example, a user might create a cell that defines \texttt{df} as a data frame (data in a table-like structure) and restrict write access to prevent other collaborators from editing the cell that declares \texttt{df}.
However, collaborators could still create a new cell that either re-declares or mutates the value of \texttt{df} and breaks downstream code that reference it.

Thus, \sys{} also introduces \emph{variable-level access control}.
Variable-level access control extends the idea of access control from cells to shared variables---authors can determine if collaborators' code can view or modify the values of runtime variables.
\sys{} tracks the runtime state of the notebook kernel and extracts the variable information.
Users can specify the access control of every variable in a side panel (Figure \ref{fig:overview}.B).
On the other collaborators' side, the protected variable is highlighted throughout the notebook.
When an individual attempts to execute a code cell a static analysis on the abstract syntax tree (AST) of the program is done to determine whether the execution would impact the value of protected variables and, if so, the execution is halted with an error.

Variable-level access control is especially beneficial for scenarios where there is a lead collaborator in charge of managing important data tables.
Setting variable-level access control can encourage collaborators to either make a copy or use parallel cell groups before they do any risky explorations.

\subsection{Parallel Cell Groups}
%
Data science work is often exploratory.
Authors might write code to explore an idea or approach.
In the context of teams, multiple team members might simultaneously work through different approaches for the same problem~\cite{wang2019data}.
In these situations, authors might want to write code that manipulates \emph{their own version of} some subset of variables in the notebook.

\sys{} thus also introduces \emph{parallel cell groups} (which we will call ``parallel cells'').
Parallel cells define a designated area where changes of the code and runtime state stay inside its own scope.
As Figure \ref{fig:overview}.C shows, users can split a regular code cell into parallel cell groups.
Collaborators can create new cell groups to branch off and explore alternatives; add multiple cells to a cell group to write larger and more complex alternative code; and work individually in each cell group.
The parallel cell groups are folded together into the same area in the notebook, helpings collaborators to maintain an overall coherent structure of the narrative.
In addition, when collaborators are settled on a solution, they can mark a cell group as ``primary'', which merges the execution result into the main runtime state.
Note that the parallel cell groups designed in \sys{} are different notions than the forked cells in \cite{weinman2021fork} in several ways.
In terms of the usage scenario, \cite{weinman2021fork} is designed for a single developer to explore alternative ideas, whereas \sys{} is designed for synchronous computational notebooks.
For the implementation, \sys{} uses a scoping mechanism instead of spawning multiple kernels, making it easier for managing different versions of the same variable.

A key difference from prior tools for branching and managing local versions \cite{kery2017variolite} is that each cell group has its own execution scope---changing the variables in one cell group would not affect others.
For example, suppose there is a parallel cell group is named \texttt{plel}, and within those cells, code creates variables named \texttt{x} and \texttt{y}.
Inside of the cell group, \texttt{x} and \texttt{y} can be referenced as usual.
Outside of the cell group, code that references variables \texttt{x} and \texttt{y} get their `old' values---whatever value they were assigned to outside of the cell group.
However, these variables can be referenced outside of the group if the user explicitly specifies which scope they want to reference.
So while \texttt{x} and \texttt{y} are not affected outside of \texttt{plel}, collaborators can refer to \texttt{\_plel.x} and \texttt{\_plel.y} to access the values that were set inside of \texttt{plel}.

Parallel cell groups allow collaborators to flexibly split the notebook for exploring alternatives.
It is particularly designed for the ``competitive authoring'' collaboration style \cite{wang2019data} where team members competitively write code for the same purpose and reach consensus when an acceptable solution is found.
This allows collaborators to work independently while making concurrent edits and executions, preventing costly mismatches between programmers' mental state and the actual state of the runtime.
It also provides collaborators with the shared context so they do not work too ``far'' away from one another, thus supporting awareness of the others' actions (e.g., working on an individual notebook for exploration).
Finally, this feature preserves the structure of the narratives by grouping and folding parallel alternatives together.

In \sys{}, parallel cell groups are represented as indented cells.
Conceptually, this matches the semantic meaning of indentation in Python (specifying the bounds of a code block and potentially creating a new scope).

\section{Evaluation}
We conducted a laboratory study on handling a scenario that involves editing conflicts with a paired collaborator.
This conflict editing scenario is synthesized from literature \cite{wang2019data} and reproduced by pairing participants with a member of the research team who plays the role of a ``clumsy collaborator''.
\subsection{Study Procedure}
The clumsy collaborator communicated with the participant using a simple chat interface we developed.
%
Participants were asked to collaborate with another individual (the clumsy collaborator).
They were informed that not all the study procedures would be explained until the end of the study, and we did not reveal the clumsy collaborator being a member of the research team.
After a brief demonstration of the RTC feature, we asked the participants to use the chat to greet the clumsy collaborator, who introduced themselves as a data science student who knew Python and regex, but was not experienced in Pandas.
Next, we explained the task.
The task was adapted from a Kaggle challenge to preprocess customer support Twitter contents and has three sub-goals: lower casing (T1), removing Twitter usernames (T2), and removing URLs (T3).
In the notebook, we also inserted sections on removing punctuation and removing frequent words after T3, with code already implemented.

We then asked the participant to work with the clumsy collaborator, who followed a script to create conflict editing scenarios, and provide hints when the participant was stuck for a certain period of time on each sub-goals.
For participants to get familiar with the RTC feature and the clumsy collaborator, we first asked the participant to work with the clumsy collaborator to solve T1, where the clumsy collaborator will not disturb the participant’s work.

Next, we asked the participant to solve T2 without the conflict editing feature.
The clumsy collaborator would propose to work in separate cells below T3, and \say{accidentally} execute a script that changes the column the participant is working on to disturb their work.
We would observe how participants reacted to the unexpected execution results.

Following a demo of \sys{}, we asked the participant to solve T3 with the conflict editing feature.
The clumsy collaborator would follow the participants' suggestion to use the notebook, and \say{accidentally} execute the script to change the dataset again.

\subsection{Results}
\subsubsection{Conflict editing is hard to notice and prevent}
After the second task, most participants (13/14) were not able to correctly find out what caused the code cell not to return the expected results until we explained it to them.
This aligned with our observations that many participants (12/14) switched their browsers to search for API documentation and did not stay on the shared notebooks all the time.
Moreover, several participants (P5, P10) did not even notice that the output was wrong.
There was an exceptional case where P9 ran the data loading cell right before executing the cell for removing the twitter username, leaving no chance for the clumsy collaborator to modify the shared variable.
P9 explained that they prefer to reload data every time before executing a new cell unless the data frame is very large.

Interestingly, although several participants (4/14) were able to recover from the issue by reloading the dataframe, they still did not identify the source of the problem.
Most participants (12/14) did not doubt their collaborators' actions or question what they did. 
Instead, they blamed themselves and looked into their own code to debug.
For example, P3 said:
\begin{quote}
    I felt like I had the correct code. But I assumed something was wrong with it. I just didn't even think that it could have been the collaborator's code.
\end{quote}


\subsubsection{Perceptions of \sys{} for preventing conflict editing}
\label{paired-session-participant-perceptions}

In T3, participants used \sys{} to work with the clumsy collaborator.
All participants (14/14) chose to create parallel cell groups and suggested the clumsy collaborator to write their code in a parallel cell.
After they finished the task, some participants (8/14) cleaned up the notebook by unindenting the parallel cell groups.
Several participants (2/14) chose to keep the clumsy collaborator's parallel cell and merge their solution into the notebook by marking their solution as main.

Overall, participants reported that they felt confident about not messing up with the shared notebook.
In particular, participants mentioned that the parallel cell groups made the shared notebook \say{neat} (P4), \say{organized} (P11), and \say{structured} (P6).
Although participants did not use the cell-level access control and variable-level access control, they described scenarios where these features could be useful.
P10 mentioned that both features could be helpful in the large classroom setting where an instructor has a sample notebook.
P5 said that variable-level access control can be useful when the cost of restarting the kernel and running previous code cells is expensive.
He described the scenario where a data science manager would not want interns to accidentally modify large-scale data tables and had to restart the kernel to recover the results.
In addition, P10 mentioned that she would use the cell-level access control on finished code cells, and use parallel cell groups on work-in-progress cells.
Noticeably, several participants mentioned that read access in cell access control was not necessary for themselves, but they could see it being used by other people.

\subsubsection{Improvement of the Parallel Cell}
Participants had several ideas on how to improve the parallel cell feature in \sys{}.
Several participants (P6, P9) mentioned adding notifications or activity histories to track if others have unindented a code cell:
\begin{quote}
    When you merge the selected tab with the main thread, that's like a commit to the main repository in GitHub. So then, you know, you need to also tell others that I have launched this, maybe a notification. I was hoping there would be some way to track that, like GitHub provides a history of commits that somebody has made to other changes.
\end{quote}
In addition, P2 asked for a merging process where she could pull cells from various fragments:
\begin{quote}
    I wish there is an option to merge different parts of the cell, like one collaborator has one cell, and then you merge the second part of another collaborator.
\end{quote}

\subsubsection{Resonate with prior experience}
The instance of editing conflict in task 2 resonated with participants' prior experience with real-time collaborative editing.
P1 mentioned a different collaborative setting in a data science classroom.
The data science classroom had around 100 students and the instructor asked everyone to join the same notebook in Deepnote.
However, the instructor asked students to not directly run code cells in the notebook.
Instead, students typed out solutions and commented at the same time.
Several participants (P9, P13) mentioned that their prior experience with shared notebooks was mostly asynchronous collaborating.
\section{Discussion and Future Work}


Our work suggests exciting opportunities for supporting collaborative editing at scale.
For example, instructors can share a collaborative notebook with a classroom of students; researchers can share a collaborative notebook with a broader audience for open collaboration; data science hobbyists can make their live streaming session more engaged by sharing the collaborative notebook session.
Future work can use mechanisms like searching, tagging, and filtering for managing parallel cells.
Our current design of cross-referencing allows participants to computationally compare versions of variables from different parallel cells, which could be improved by integrating visualization techniques to compare data changes \cite{wang2022diff}, or clustering techniques to explore variance \cite{glassman2015overcode}.
We are also interested in incorporating domain-specific features, such as testing students' code cells with peer-written test cases \cite{wang2021puzzleme}.


Another key area for future work is in improving users' awareness of collaborators' activity.
For example, parallel cell groups allow multiple users to work on different versions of the same document concurrently, but makes it difficult for users to see each other's cursor movements or edits.
In addition, the design of \sys{} brings up the unique challenges in helping collaborators track and forage editing history.
With non-linear notebook structures, it is worth exploring how notebook history foraging designs \cite{kery2019towards} can be extended to support the awareness of complex cell editing.

\section{Conclusion}
Real-time collaborative editing in computational notebooks requires strategic coordination between collaborators.
We investigated common obstacles in real-time notebook editing and proposed a set of access control mechanisms to support conflict-free editing: \emph{cell-level access control} (which restricts collaborators' ability to see or edit cells), \emph{variable-level access control} (which protects runtime variables from being referenced or modified), and \emph{parallel cell groups} (which allow collaborators to work in their own space while staying connected to the larger notebook).
As we found in our user studies with \sys{}, these features can improve collaboration within data science teams.

\bibliographystyle{ACM-Reference-Format}
\bibliography{main}



\end{document}